\documentclass[namedreferences]{solarphysics}

\usepackage[hyperref,optionalrh]{spr-sola-addons} 
\usepackage{graphicx}        
\usepackage{color}           
\usepackage{breakurl}        
\usepackage{multirow}	
\usepackage{lipsum}
\usepackage{enumerate}

\begin{document}

\begin{article}

\begin{opening}

\title{A statistical correlation of sunquakes based on their seismic, white light, and X-ray emission }

\author{J.C.~\surname{Buitrago-Casas}$^{1,2}$\sep
        J.C.~\surname{Mart\'inez Oliveros}$^{1}$\sep
        C.~\surname{Lindsey}$^{3}$\sep
        B.~\surname{Calvo-Mozo}$^{2}$\sep
        S.~\surname{Krucker}$^{1,4}$\sep
        L.~\surname{Glesener}$^{1}$\sep
        S.~\surname{Zharkov}$^{5,6}$
       }
\runningauthor{J.C. Buitrago-Casas et al.}
\runningtitle{Seismic, white light and X-ray emission relationship of solar flares}

   \institute{$^{1}$ Space Sciences Laboratory, University of California, Berkeley, CA, USA
                     email: \url{milo@ssl.berkeley.edu} email: \url{oliveros@ssl.berkeley.edu} email: \url{krucker@ssl.berkeley.edu} email: \url{glesener@ssl.berkeley.edu} \\ 
                  $^{2}$ Observatorio Astron\'omico Nacional, Universidad Nacional de Colombia, Bogot\'a, Colombia 
                  email: \url{bcalvom@unal.edu.co}\\
                  $^{3}$ Colorado Research Associates Division - North West Research Associates, Boulder, CO, USA 
                     email: \url{clindsey@cora.nwra.com} \\
                  $^{4}$ University of Applied Sciences Northwestern Switzerland\\
                  $^{5}$ Department of Physics and Mathematics, University of Hull, Cottingham Road, Kingston-upon-Hull, HU6 7RX, UK email: \url{S.Zharkov@hull.ac.uk}\\
                  $^{6}$ Mullard Space Science Laboratory, University College London, Holmbury St. Mary, Dorking RH5 6NT, UK
             }

\begin{abstract}
Several mechanisms have been proposed to explain the transient seismic emission, i.e., ``sunquakes,'' from some solar flares. Some theories associate high-energy electrons and/or white-light emission with sunquakes. High-energy charged particles and their subsequent heating of the photosphere and/or chromosphere could induce acoustic waves in the solar interior. We carried out a correlative study of solar flares with emission in hard-X rays (HXRs), enhanced continuum emission at 6173\AA, and transient seismic emission. We selected those flares observed by {\it RHESSI} ({\it Reuven Ramaty High Energy Solar Spectroscopic Imager}) with a considerable flux above 50 keV between January 1, 2010 and June 26, 2014. We then used data from the Helioseismic and Magnetic Imager onboard the {\it Solar Dynamic Observatory} (SDO/HMI) to search for excess visible continuum emission and new sunquakes not previously reported. We found a total of 18 sunquakes out of 75 investigated. All of the sunquakes were associated with a enhancement of the visible continuum during the flare time. Finally, we calculated a coefficient of correlation for a set of dichotomic variables related to these observations. We found a strong correlation between two of the standard helioseismic detection techniques, and between sunquakes and visible continuum enhancements. We discuss the phenomenological connectivity between these physical quantities and the observational difficulties of detecting seismic signals and excess continuum radiation.
\end{abstract}
\end{opening}

\section{Introduction}
     \label{S-Introduction} 

Solar flares are impulsive events that transform magnetic energy mainly into accelerated charged particles, plasma waves, radiation, and heating the solar atmosphere. Some energetic flares produce acoustic disturbances observed as expanding ripples in the photosphere. This phenomenon, known as a {\it sunquake}, was first suggested by \cite{wolf1972} and, a couple of decades after, observed for the first time by \cite{kosovichev1998}. Although several authors have proposed different hypotheses to explain the generation of acoustic sources, currently there is no agreement about the exact mechanism.\\

Leading hypotheses for sunquake generation include: (i) Direct impact of high-energy charged particles with the lower atmosphere layers, heating them; (ii) back-warming of the photosphere by radiation from an overlying heated chromosphere, and (iii) a transient Lorentz force.
\begin{enumerate}[i]
\item In this idea a beam of charged particles (electrons, protons and/or ions) accelerated in the solar corona deposit energy in the chromosphere and/or the photosphere. This results in an explosive evaporation that creates a downward-propagating shock wave, causing the seismic response \citep{ZZ2007,Z2008}. A crucial quality for this theory is the penetration power of charged particles \citep{donea2005,kosovichev2007,ZZ2008}. Depending on solar atmospheric conditions and the particle beam characteristics, particles and/or shocks may or may not reach deeper layers of the Sun, causing a sunquake.

\item A second hypothesis suggests that the generation of sunquakes is related to sudden heating of the solar photosphere during the impulsive phase of a flare. Some authors \citep{donea2006b,donea2006,pedram2012} have found a temporal and spatial correlation between seismic signatures and an intensity augmentation of visible-continuum emission.  These results support the back-warming model as a plausible mechanism for sunquake generation. In this hypothesis a particle beam precipitating into the chromosphere produces an increase in the Balmer and Paschen emission edges through collisional heating and ionization of the chromosphere. Then, the photosphere is heated by absorbing most of the portion of this radiation that emanates downward from the chromosphere. If the whole process occurs quickly enough, the result will be an increase in the total local photospheric pressure which, in turn, will drive a compact acoustic disturbance into the sub-photosphere \citep{Z1993}.

\item A third possible explanation posits a transient Lorentz force as a driver for sunquakes. In accordance with \cite{zk2002}, reversible transient changes in the solar flare magnetic field apply a transient force to the photosphere that drives a seismic transient. \cite{h2008,mo2009} and \cite{f2012} theoretically and observationally assessed this hypothesis and found it as a plausible mechanism to generate sunquakes. Most recently, some  studies have assessed changes in the magnetic energy \citep{alvarado2012} and have analyzed the structure of the magnetic field \citep{xu2014,liu2014} in the context of seismic transients driven by Lorentz forces.
\end{enumerate}

Other mechanisms proposed to explain sunquake generation include heavy ions, pions, gamma rays, and sudden electric field changes \citep{ZZ2007,zharkova2008,sharykin2014}.\\

In this work, we selected a sample of flares in solar cycle 24 with emission in hard X-rays (HXR) observed by the {\it Reuven Ramaty High Energy Solar Spectroscopic Imager} ({\it RHESSI}), indicationg the presence of accelerated electrons. We used data from the Helioseismic and Magnetic Imager on board  the {\it Solar Dynamic Observatory} (SDO/HMI) to identify visible-continuum and Doppler features for every flare in the sample. We applied standard helioseismic techniques to determine whether a sunquake obviously occurred in each event. We found several new sunquakes that have not been reported before. The analysis is explained in section \ref{S-Observations}. The main results of the survey, and a statistical correlation analysis, are presented in section \ref{S-Results}. Finally, in section \ref{S-Conclusions} we summarize our main results and discuss future work.

\section{Observations} 
      \label{S-Observations}  

So far, sunquakes are believed to be a solar flare related phenomenon. Although flares are commonly observed events, sunquakes are less so. This may be due to limitations in the instruments and techniques used for their detection. That is why a statistical study of sunquakes is advantageous to identify a suitable indicator that readily allows us to find seismic events \citep{b2008,pedram2012}.\\

\subsection{Event selection}
      \label{S-Events-selection} 

Sunquakes reported so far have typically been associated with large and medium GOES-class solar flares (e.g., \cite{mo2008}). However, powerful flares with no detectable seismic response are also common, e.g. \cite{liu2014}. This suggests that the GOES soft X-ray classification is not an adequate proxy for acoustic signatures. Instead, we used HXRs, because of their direct relation with electron acceleration and their possible relation with the generation of a sunquake in accordance with hypotheses {\it i} and {\it ii} mentioned above.  For this reason we used HXR measurements from {\it RHESSI} to choose our events.\\

{\it RHESSI} is a spacecraft launched on February 2002 that has been continuously observing X-rays from the Sun \citep{l2002}. For our sample we chose those flares for which RHESSI registered a clear observation of X-rays above 50 keV during most of solar cycle 24.\\

The events for our sample were taken from the {\it RHESSI} flare list\footnote{The automatically generated solar flare list observed by {\it RHESSI} can be found at \url{http://hesperia.gsfc.nasa.gov/hessidata/dbase/hessi_flare_list.txt}} avoiding those flares that took place within 100 arcsec of the Sun's limb, because of observational difficulties in detecting sunquakes at those locations. Thus, our sample consists of 75 solar flares from which X-rays above 50 keV emanated in the period between January 1, 2010 and June 26, 2014 (the start of the fourth {\it RHESSI} anneal). This set of events included 10 X-, 46 M-, and 19 C-class flares.\\

Using standard helioseismic techniques, we carried out a survey to determine the existence, or not, of a sunquake.  We also used HMI ``continuum'' maps to measure concurrent enhancements in continuum emission in the neighborhood of the $6173.34${\AA} Fe I absorption line emanating from the seismic source region.

\begin{figure}    
\centerline{\includegraphics[width=1.0\textwidth,clip=]{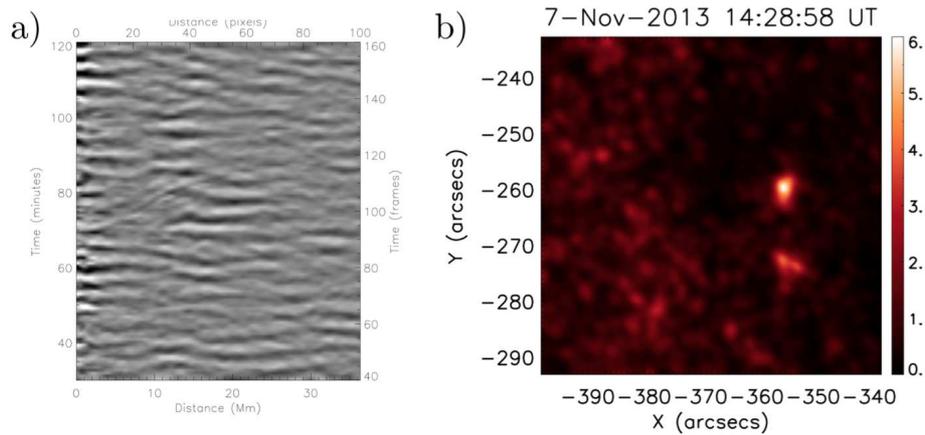}}
\caption{Helioseismic detection techniques applied to a sunquake on November 7, 2013 at 14:28 UT. {\bf a)} {\it Time-distance plot} generated on a temporal sequence of consecutive Doppler differences with a pupil centered at $x_c=-363$ arcsec and $y_c=-263$ arcsec. In this case the ``sweep angle'' (see text) was taken over the first and second quadrants. In this plot, $t=60$ minutes corresponds to 14:28 UT, the time for the seismic disturbance. The signature of outgoing ripples emanating from the center reference of the plot is a faint, downwardly curved ridge connecting the points (5~Mm, 75~s) and (18~Mm, 85~s) in the plot. {\bf (b)} {\it Egression acoustic map} where it is easy to identify two acoustic kernels. Color scale is normalized to the egression background.}
\label{F-helioseismic}
\end{figure}

\subsection{Helioseismic analysis}
      \label{S-Helioseismic-analysis} 

Sunquakes are difficult events to observe. Their signatures are usually weak compared to the ambient acoustic noise observed in the photosphere. Despite this difficulty, some sensitive methods have been developed for their detection. The time-distance plots \citep{kosovichev1998} and helioseismic holography \citep{lb1997} are the techniques usually used for this purpose. Both techniques are applied to a time series of HMI Dopplergrams.\\

In this paper, we used Dopplergrams from the SDO/HMI. This instrument has 6 tuned filters around the $6173.34${\AA} Fe I absorption line, with a $76\pm10${m\AA} bandwidth. Every Dopplergram is computed on the ground using twelve filtergrams of the Sun (with a cadence of 3.75 seconds) obtained with sequences of tuned wavelengths and polarization in a period of 45 seconds as described by \cite{hmi}. The HMI gives us motion maps of the Sun with a pixel size of 0.5 arcsec and a cadence of 45 seconds.\\

We now briefly describe the use of time-distance plots and helioseismic holography to identify sunquakes.

\subsubsection{Time distance plots}
      \label{S-Time-distance-plot} 

This technique maps the helioseismic signal averaged over thin annuli centered upon a selected reference location, as a function of time, along the vertical axis, and the radii of the respective annuli, along the horizontal, derived from a temporal sequence of Doppler differences.\\

For all of the sunquakes we observed a clear pattern of ripples propagating outward from the seismic source on Doppler difference movies. Then, we constructed time-distance plots for those flares in which we saw clear ripples propagating outward the photospheric impact location as a way to display our finding. In the time-distance plots, seismic signatures associated with a sunquake form a  curved ridge traversing the plot, as shown, for example, in Figure \ref{F-helioseismic}a. This technique is sensitive to the the reference location which should coincide with the seismic source region.  The signature can be optimized choosing an azimuthal angle range, the ``sweep angle,'' over which to compute the annular averages.  In our case, we centered the annuli at the location where we saw a local Doppler transient signature (photospheric flare impact) in the impulsive phase of the flare, and we limited the angular range to the directions along which we saw strong ripples propagating.  As an outcome of this procedure, we found a total of 18 seismically active flares in our 75-events sample (Table \ref{T-sunquakes})\\

These techniques have some strengths and weaknesses. This is a simple and easy method to implement that does not use strong assumptions (such as an acoustic model) other than the primary location of the source, and can be directly applied to a set of Doppler difference maps.  But, due to the photospheric noise, most of the time it is difficult to distinguish a seismic signature associated  with a sunquake.  Also, the effectiveness of this technique depends strongly on the reference location and sweep angle, which currently are judged somewhat subjectively.

\subsubsection{Helioseismic holography}
      \label{S-Helioseismic-holography} 

Computational seismic holography was introduced by \cite{lb1990,lb1997} as a diagnostic whose essential object is optimal local discrimination of compact sources and anomalies in the solar acoustic environment.  This technique applies basic principles of wave optics to helioseismic signatures at the Sun's surface to coherently image the source power distribution of the waves. Helioseismic holography was initially conceived as a means of imaging acoustic sources and anomalies in the solar interior.  However, most of the acoustic radiation released by flares penetrates only a few tens of Mm before it refracts back to the surface, hence the ripples described in \S \ref{S-Introduction}.  Seismic holography, then, can image acoustic sources at the photosphere, sampling said ripples as they pass through an annular ``pupil'' surrounding the source region and extending tens of Mm from it in radius.\\

This diagnostic has proved to be a powerful detection tool for sunquakes \citep{z2013b}.  Figure \ref{F-helioseismic}-b shows a source distribution map generated by the application of seismic holography to HMI Doppler observations in the 5--7-mHz acoustic spectrum (commonly chosen spectral range) for the flare that occurred on November 7, 2013 at 14:29 UT.  Although there has not been reported a one-to-one comparison survey so far, helioseismic holography has generally agreed with the results shown by the time-distance plots \citep{zg2013,zgm2013}.\\

For every flare in our sample we generated its respective 5 - 7 mHz source distribution map to identify any acoustic signatures around the flare time and location (again, we took times and positions from the RHESSI list). We found 18 events that showed at least one sharp kernel in the source distribution map with a significant excess of seismic emission easily recognizable by eye. These 18 events coincide with the same 18 sunquakes found using the time-distance plots.

\begin{figure}
\centerline{\includegraphics[width=1.0\textwidth,clip=]{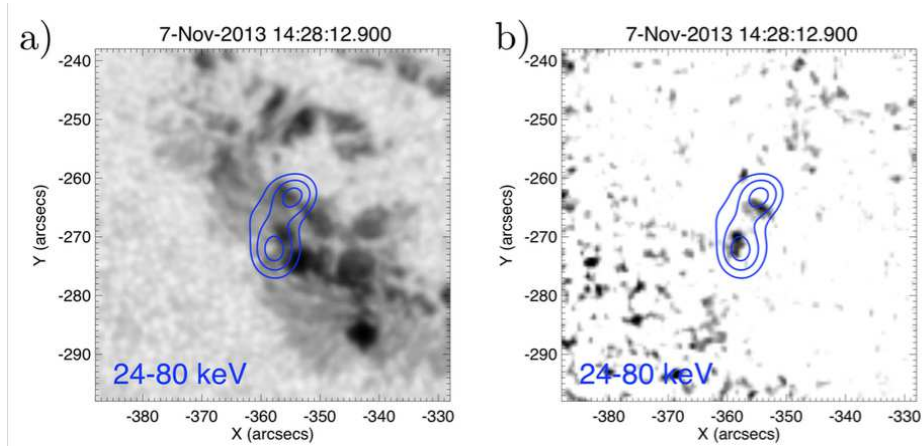}}
\caption{White Light and HXR emission plots. {\bf (a)} HMI intensity map of the active region 11890 for the flare on November 7, 2013 at 14:28 UT. {\bf (b)} HMI intensity differences with the grey color scale inverted for the same time. Both plots show an over-plotted set of blue contours that correspond to the HXRs observed by RHESSI. This RHESSI image was generated using the standard CLEAN algorithm and subcollimators  3 - 5 integrating between 14:27:55 and 14:31:13 UT.}
\label{F-wl}
\end{figure}

\begin{figure}    
\centerline{\includegraphics[width=0.9\textwidth,clip=]{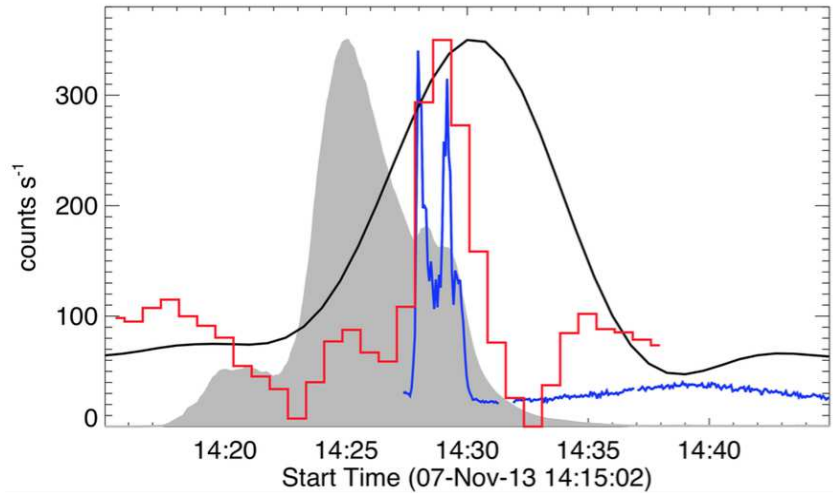}}
\caption{Time profile at different wavelengths for the flare that occurred on November 7, 2013 at 14:28 UT. Because the goal of this plot is to show temporal coincidences, the y-axis is in arbitrary units. The grey filled background is the GOES curve. The blue line is the HXR emission as observed by RHESSI in the 24-80 keV range. The red line is the visible-continuum intensity profile based on HMI data enclosed in a mask around the transient signal location. The black line represents the acoustic source power time profile in the 5 to 7 mHz range. A temporal coincidence between  the HXR and the excess visible-continuum emission is clearly observed. Also, the temporal behaviour of the acoustic source power agrees with what is expected from an acoustically active flare, that is, the visible-continuum and HXR peaks are within the FWHM of the black line.}
\label{F-time}
\end{figure}

\subsection{Identification of visible continuum emission}
      \label{S-White-light-emission-analysis} 

Brightening of visible continuum emission, if extended over a broad spectrum, indicates an intense energy release in solar flares. While these sources are intense, our ability to discriminate these sources is limited by noise generated by photospheric convection motions and acoustic oscillations \citep{h2010}. To assess excess continuum emission in each flare of our sample, we used standard HMI level-1p ``continuum intensity maps'' published by the Stanford Joint Science Operations Center (JSOC) characterizing the continuum in which the $6173.34$\AA \,Fe I absorption line is embedded.  This photospheric line was chosen for SDO/HMI because it is found within a clean continuum with no blends and no near-by lines \citep{stenflo1977,norton2006}.  We made a temporal sequence of SDO/HMI continuum intensity maps of consecutive differences (running differences) for every flare in our list. The sequence includes times 2 hours before and after the flare time, and were centered on the flare location (according to the RHESSI flare-list) with a FOV of $300\times300$ arcsec$^2$. \\

In some of the temporal sequences it was easy to observe a clear augmentation, by eye, of the SDO/HMI intensity maps corresponding to the flare time and location (see Figure  \ref{F-wl} a and b). We call flares that exhibited this augmentation ``white-light flares,'' after flares known to exhibit very similar enhancements over a broad spectrum, bearing in mind that the excess registered by HMI is only in a narrow band of the spectrum.\\




In general, the HXR emission temporally coincides with the white light continuum enhancement, the Doppler transient signature, and the appearance of acoustic kernels in the seismic source distribution maps. As an example, Figure \ref{F-time} shows the time profile of different wavelengths and their relationship to the sunquake that occurred on November 7, 2013 at 14:28 UT. That plot certainly evidences a temporal coincidence between the HXR and the excess visible-continuum emission. Also, the temporal behaviour of the acoustic source power agrees with what is expected from an acoustically active flare, that is, the visible-continuum and HXR peaks are within the FWHM of the black line.

\section{Results}
      \label{S-Results}

\begin{table}[ht!]
\caption{Times and locations of the seismically active solar flares found in this survey. Some of these have been reported by other authors before, e.g.,  \cite{kosovichev2011,k2014,zg2013,sharykin2014}. Peak times and positions were taken from the RHESSI flare list.}
\label{T-sunquakes}
\begin{tabular}{llllll}\hline
\# & Date       & Peak time    & X (arcsec)    & Y (arcsec)   & GOES \\\hline
1 & 2011-02-15 & 1:55:30  & 205  & -222 & X2.2 \\
2 & 2011-07-30 & 2:09:10  & -526 & 170  & M9.3 \\
3 & 2011-09-26 & 5:07:58  & -519 & 116  & M4.0 \\
4 & 2012-03-09 & 3:34:52  & 0    & 389  & M6.3 \\
5 & 2012-05-10 & 4:17:42  & -364 & 259  & M5.7 \\
6 & 2012-07-04 & 9:54:26  & 289  & -343 & M5.3 \\
7 & 2012-07-05 & 3:35:50  & 417  & -338 & M4.7 \\
8 & 2012-07-05 & 11:44:14 & 495  & -332 & M6.1 \\
9 & 2012-07-06 & 1:39:38  & 585  & -322 & M2.9 \\
10 & 2012-10-23 & 3:17:22  & -795 & -272 & X1.8 \\
11 & 2013-02-17 & 15:50:22 & -338 & 307  & M1.9 \\
12 & 2013-07-08 & 1:22:10  & 75   & -217 & C9.7 \\
13 & 2013-11-06 & 13:48:42 & -549 & -267 & M3.8 \\
14 & 2013-11-07 & 3:39:38  & -450 & -272 & M2.3 \\
15 & 2013-11-07 & 14:28:22 & -363 & -263 & M2.4 \\
16 & 2014-01-07 & 10:16:14 & -228 & -168 & M7.2 \\
17 & 2014-02-02 & 6:33:54  & -300 & 314  & M2.6 \\
18 & 2014-02-07 & 10:28:38 & 764  & 270  & M1.9 \\\hline
\end{tabular}
\end{table}

Besides the criterion for 50~keV X-rays, we studied three other aspects of each flare in our sample of 75 events: We investigated the existence of a transient Doppler signal,  an enhancement in the continuum emission, and the seismic emission, as indicated by time-distance plots and helioseismic holography. As a result, we found 28 white light flares, 18 sunquakes through a helioseismic holography and time-distance detections, and 41 flares with a clear Doppler transient signature. Information about the 18 seismically active flares found using both helioseismic techniques is showed in Table \ref{T-sunquakes}.\\

\subsection{Dichotomic Correlations} 
      \label{S-Correlation} 

A standard statistical technique to measure the correlation between two dichotomic variables\footnote{A dichotomic variable is a variable that can take one of two possible values \citep{cramer1999}.} is  the {\it mean square contingency coefficient} \citep{cramer1999}. If $X_1$ and $X_2$ are two dichotomic variables with {\it yes} and {\it no} as possible values, the mean square contingency coefficient, $\phi$, between these two variables is written as\\

\vspace{0.5cm}

  \begin{minipage}[c]{0.4\textwidth}
    \centering
	\begin{tabular}{cc|c|c|c|c|l}
	\cline{3-4}
	& & \multicolumn{2}{ c| }{$X_1$} \\ \cline{3-4}
	& & no & yes  \\ \cline{1-4}
	\multicolumn{1}{ |c  }{\multirow{2}{*}{$X_2$} } &
	\multicolumn{1}{ |c| }{yes} & $a$ & $b$   \\ \cline{2-4}
	\multicolumn{1}{ |c  }{}                        &
	\multicolumn{1}{ |c| }{no} & $c$ & $d$    \\ \cline{1-4}
	\end{tabular}
  \end{minipage}
  \begin{minipage}[c]{0.5\textwidth}
    \centering	
  	\begin{equation}
	\phi = \frac{cb - ad}{\sqrt{(a+b)(c+d)(a+c)(b+d)}}
	\end{equation}
  \end{minipage}

\vspace{0.5cm}

This coefficient can take on values between $-1$ and $1$. The sign indicates the direction of the correlation and the number indicates the strength. $\phi=0$  indicates the absence of any  correlation between the two variables.\\

We selected the observation of a Doppler transient signature, the existence of a visible continuum enhancement, and the existence of a sunquake using the time-distance plot and helioseismic holography as our set of four dichotomic variables. Results for the mean square contingency coefficient calculated over each possible pair of our four variables are shown in Tables \ref{T-seismic} and \ref{T-wl}.\\

All sunquakes we found were related to a white light flare and every white light flare was associated with a Doppler transient signature. Nevertheless, not all flares with a Doppler transient showed an enhancement in the visible continuum, and not all white light flares had an associated sunquake. \\

From Table \ref{T-seismic} we find a strong correlation between the two helioseismic techniques that we used: The time-distance plot and helioseismic holography. This result is expected since both techniques are based on the same data: Doppler observations from the SDO/HMI.\\

A moderately strong correlation was found between sunquakes and Doppler transient signatures. It is related to the fact that all of the seismically active flares also showed a Doppler transient. However, this correlation factor is not exactly unity because not all of the flares that showed a Doppler transient hosted a sunquake. As seen from Table \ref{T-seismic}, this result does not depend on the helioseismic technique. 

  \begin{table}\caption{Dichotomic correlations among Doppler transient, time-distance plot, and helioseismic holography signatures.}
  \label{T-seismic}

	\begin{tabular}{p{.8cm}p{.8cm}|p{1cm}|p{.8cm}|p{.8cm}|}
	\cline{3-5}
	& & \multicolumn{3}{ c| }{TD-plot} \\ \cline{3-5}
	& & \multicolumn{1}{ |c| }{no} & \multicolumn{1}{ |c| }{yes} & \multicolumn{1}{ |c| }{total}  \\ \cline{1-5}
	\multicolumn{1}{ |c  }{\multirow{2}{*}{Holograp.} } &
	\multicolumn{1}{ |c| }{yes} & 0 & 18 & 18  \\ \cline{2-5}
	\multicolumn{1}{ |c  }{}  & 
	\multicolumn{1}{ |c| }{no} & 57 & 0 & 57  \\ \cline{1-5}
	& \multicolumn{1}{ |c| }{total} & 57 & 18 & 75 \\ \cline{2-5}
	\end{tabular}
$\phi = 1.00$

\vspace{0.5cm}

	\begin{tabular}{p{.8cm}p{.8cm}|p{1cm}|p{.8cm}|p{.8cm}|}
	\cline{3-5}
	& & \multicolumn{3}{ c| }{Doppler-transient} \\ \cline{3-5}
	& & \multicolumn{1}{ |c| }{no} & \multicolumn{1}{ |c| }{yes} & \multicolumn{1}{ |c| }{total}  \\ \cline{1-5}
	\multicolumn{1}{ |c  }{TD-plot \&} &
	\multicolumn{1}{ |c| }{yes} & 0 & 18 & 18  \\ \cline{2-5}
	\multicolumn{1}{ |c  }{Holograp.}  & 
	\multicolumn{1}{ |c| }{no} & 34 & 23 & 57  \\ \cline{1-5}
	& \multicolumn{1}{ |c| }{total} & 34 & 41 & 75 \\ \cline{2-5}
	\end{tabular}
$\phi = 0.51$
\end{table}

  \begin{table}\caption{Dichotomic correlations between the visible continuum enhancements and each of the acoustic variables.}
  \label{T-wl}
	\begin{tabular}{p{.8cm}p{.8cm}|p{1cm}|p{.8cm}|p{.8cm}|}
	\cline{3-5}
	& & \multicolumn{3}{ c| }{Doppler-transient} \\ \cline{3-5}
	& & \multicolumn{1}{ |c| }{no} & \multicolumn{1}{ |c| }{yes} & \multicolumn{1}{ |c| }{total}  \\ \cline{1-5}
	\multicolumn{1}{ |c  }{\multirow{2}{*}{White-light} } &
	\multicolumn{1}{ |c| }{yes} & 0 & 28 & 28  \\ \cline{2-5}
	\multicolumn{1}{ |c  }{}  & 
	\multicolumn{1}{ |c| }{no} & 34 & 13 & 47  \\ \cline{1-5}
	& \multicolumn{1}{ |c| }{total} & 34 & 41 & 75 \\ \cline{2-5}
	\end{tabular}
$\phi = 0.70$

\vspace{0.5cm}
	
	\begin{tabular}{p{.8cm}p{.8cm}|p{1cm}|p{.8cm}|p{.8cm}|}
	\cline{3-5}
	& & \multicolumn{3}{ c| }{TD-plot \& Holograp.} \\ \cline{3-5}
	& & \multicolumn{1}{ |c| }{no} & \multicolumn{1}{ |c| }{yes} & \multicolumn{1}{ |c| }{total}  \\ \cline{1-5}
	\multicolumn{1}{ |c  }{\multirow{2}{*}{White-light} } &
	\multicolumn{1}{ |c| }{yes} & 10 & 18 & 28  \\ \cline{2-5}
	\multicolumn{1}{ |c  }{}  & 
	\multicolumn{1}{ |c| }{no} & 47 & 0 & 47  \\ \cline{1-5}
	& \multicolumn{1}{ |c| }{total} & 59 & 16 & 75 \\ \cline{2-5}
	\end{tabular}	
$\phi = 0.73$

\end{table}

\section{Discussion and Conclusions} 
      \label{S-Conclusions} 

We have identified sunquakes and visible-continuum enhancements from a sample of 75 flares that RHESSI observations show to have HXR emission above 50 keV, indicating the presence of energetic electrons. Using helioseismic holography and time-distance plots, we identified a total of 18 sunquakes, several of which have not been reported before. We then carried out a statistical study to analyze possible correlations between acoustically active flares and continuum emission, expressing those observations as dichotomic variables.\\

Key results include:

\begin{itemize}
\item Using time-distance plots and helioseismic holography we identified 18 seismically active events (sunquakes).
\item 41 flares display a clear Doppler transient signal during the flare impulsive phase that is easy to distinguish by eye on the Doppler difference maps. 
\item Not every flare with a Doppler transient signature was a white light flare. 13 of the flares with a Doppler transient did not show a significant visible-continuum enhancement.
\item Not every flare with a Doppler transient signature hosted a sunquake, but every sunquake found in this work occurred in a flare with an easily detected Doppler transient signature.
\item One third of the selected HXR flares showed a clear visible-continuum enhancement observable with SDO/HMI.
\item Every sunquake found in this survey showed a visible continuum enhancement, but not every white light flare was a seismically active event.
\end{itemize}

We found a perfect correlation between the two helioseismic detection techniques that we used, with $\phi = 1.0$.  Helioseismic holography and time-distance plots derive from the same set of data: Doppler difference maps. Thus, the dynamical range, quality, spatial resolution and cadence of the original data affects both methods equally. Because both techniques also used an integration over annuli to find sunquakes, it is natural that they yield basically the same outcomes. This result implies that helioseismic holography and time-distance plots are complementary when they are used as sunquake detectors.\\ 

Caution is necessary when analyzing our results with respect to Doppler transient signatures. Because Dopplergrams based on SDO/HMI observations are generated through a set of filtergrams, they can contain some artifacts, as described by \cite{mo2014} and \cite{norton2006}. This situation worsens when high field strengths at a particular location on the solar disk combine with high velocities. This moves the spectral line beyond its effective sampling range. The higher $g_{eff}$ of Fe I means that its useful range of velocity values in regions of strong magnetic field is small. Then, because Doppler transient signatures usually occur at locations with strong magnetic fields, they are not always reliable. Also, despite the fact that the 6173 {\AA} Fe I line is a mostly isolated line in a quiet Sun environment, it is possible that during a flare this line could become contaminated by blends or molecular absorptions.\\

This study compiles instances of an appreciable excess of visible-continuum emission from the acoustic source regions as a clue to the causal relationship between back-warming and transient seismic emission from flares. We find an appreciable excess of visible-continuum radiation from all of the acoustically active flares in the sample. In the simple model proposed by \cite{ld2008}, which supposes that back-warming is the sole contributor to transient seismic emission, we expect a fairly tight, quantitative relationship between excess transient visible-continuum emission and transient seismic emission. Flares listed in Table \ref{T-sunquakes} are a very appropriate sample on which to test this theory.\\


Now that we have identified a coherent sample of 18 sunquakes, it is desirable to carry out further, more detailed work on the relationship between HXR emission, visible continuum enhancements and seismic source locations. Additionally, a detailed temporal analysis, as well as a study of the energy associated with each phenomena, are necessary to deepen our understanding of sunquake mechanisms.

\begin{acks}
We thank Sandra M. Buitrago-Casas for her valuable comments during the writing of this article. The Berkeley group was supported by NASA under contract NAS-5-98033 for RHESSI. This paper was written based on fruitful discussions at the Leverhulme Flare Seismology Workshop held at the Mullard Space Science Laboratory (MSSL), 9-12 September 2013. Authors who attended the workshop thank the Leverhulme hosts for their hospitality during the visit to MSSL.
\end{acks}


\newpage
\bibliographystyle{spr-mp-sola}

\tracingmacros=2
\bibliography{sola_bibliography_buitrago}  

\end{article} 

\end{document}